\documentclass[aps,prl,twocolumn,superscriptaddress,nofootinbib,amssymb]{revtex4-1}

\usepackage{graphicx}
\usepackage{amsmath}
\usepackage{times}

\usepackage[usenames]{color}
\definecolor{comm}{cmyk}{0, 0.7, 0.5, 0} 

\definecolor{norm}{cmyk}{0,0,0,1}

\definecolor{luca}{cmyk}{1,0,1,0}

\begin{document}

\title{Tunable polaronic conduction in anatase TiO$_2$}

\author{S. Moser}\affiliation{Advanced Light Source (ALS), Berkeley, California 94720, USA}\affiliation{Ecole Polytechnique F\'ed\'erale de Lausanne (EPFL), Institut de Physique des Nanostructures, CH-1015 Lausanne, Switzerland}
\author{L. Moreschini}\affiliation{Advanced Light Source (ALS), Berkeley, California 94720, USA}
\author{J. Ja\'{c}imovi\'{c}}\affiliation{Ecole Polytechnique F\'ed\'erale de Lausanne (EPFL), Institut de Physique des Nanostructures, CH-1015 Lausanne, Switzerland}
\author{O. S. Bari\v{s}i\'{c}}\affiliation{Institute of Physics, Bijeni\v{c}ka c. 46, HR-10000 Zagreb, Croatia}
\author{H. Berger}\affiliation{Ecole Polytechnique F\'ed\'erale de Lausanne (EPFL), Institut de Physique des Nanostructures, CH-1015 Lausanne, Switzerland}
\author{A. Magrez}\affiliation{Ecole Polytechnique F\'ed\'erale de Lausanne (EPFL), Institut de Physique des Nanostructures, CH-1015 Lausanne, Switzerland}
\author{Y. J. Chang}\affiliation{Advanced Light Source (ALS), Berkeley, California 94720, USA}\affiliation{Department of Physics, University of Seoul, Seoul, 130-743, Korea}
\author{K. S. Kim}\affiliation{Advanced Light Source (ALS), Berkeley, California 94720, USA}
\author{A. Bostwick}\affiliation{Advanced Light Source (ALS), Berkeley, California 94720, USA}
\author{E. Rotenberg}\affiliation{Advanced Light Source (ALS), Berkeley, California 94720, USA}
\author{L. Forr\'{o}}\affiliation{Ecole Polytechnique F\'ed\'erale de Lausanne (EPFL), Institut de Physique des Nanostructures, CH-1015 Lausanne, Switzerland}
\author{M. Grioni}\affiliation{Ecole Polytechnique F\'ed\'erale de Lausanne (EPFL), Institut de Physique des Nanostructures, CH-1015 Lausanne, Switzerland}

\date{\today}

\begin{abstract}
Oxygen vacancies created in anatase TiO$_2$ by UV photons (80 -- 130 eV) provide an effective electron-doping mechanism and induce a hitherto unobserved dispersive metallic state. Angle resolved photoemission (ARPES) reveals that the quasiparticles are large polarons. These results indicate that anatase can be tuned from an insulator to a polaron gas to a weakly correlated metal as a function of doping and clarify the nature of conductivity in this material.
\end{abstract}

\maketitle

The anatase structural phase of titanium dioxide (TiO$_2$) can be the key element in novel applications. Whereas extensive work has been focused on its famous photocatalytic behavior \cite{Asahi2001,He2009,FUJISHIMA1972}, more and more proposed devices, such as memristors \cite{Strukov2009}, spintronic devices \cite{Toyosaki2004}, and photovoltaic cells \cite{OREGAN1991,Kuang2008,Varghese2009}, rely on its less well-known electronic properties. In particular, anatase has been recently suggested as a candidate for replacing the In-based technology for transparent conducting oxides \cite{Furubayashi2005} in a wide range of applications from solar cell elements, to light-emitting devices, to flat panels, to touch-screen controls\cite{Gordon2000}. The crucial quantity for the figure of merit in these devices is conductivity, and it is therefore of major interest to understand and control the electronic properties of pristine and doped anatase.

Stoichiometric anatase is an insulator with a $3.2$~eV band gap \cite{TANG1993} but oxygen vacancies, typically present with concentrations in the 10$^{17}$~cm$^{-3}$ range \cite{FORRO1994,Kitada2011}, create a shallow donor level $\sim$10~meV  below the conduction band (CB) \cite{Jacimovic2012}. Since large single crystals became available for transport studies, a better insight has been gained on the influence of these donors on the electronic response of anatase. Above  $\sim$60~K,  the electrons thermally excited into the CB give rise to metallic-like transport.  At lower temperatures, the anomalous increase of resistivity indicates that the charge carriers are not bare electrons but polarons \cite{Jacimovic2012}, \textit{i.e.}, electrons coherently coupled to a lattice distorsion induced by the Coulomb interaction.
Understanding the properties of such composite particles in anatase is important to better engineer the material for targeted applications, where the low electron mobility often represents the overall performance bottleneck. We will also demonstrate that, from the point of view of fundamental physics, anatase represents an excellent model compound to study the behavior of the ``rare'' large polaron quasiparticles (QPs), intermediate between localized small polarons and free electrons.

We performed ARPES measurements on TiO$_2$ single crystals (Fig.~1(a)) and thin films grown \textit{in situ} on insulating  LaAlO$_3$  and conducting Nb-doped  SrTiO$_3$  substrates. Clean $(001)$ surfaces were prepared as described in Suppl. Inf. The results presented  have been obtained consistently both for single crystals and thin films,  and therefore reflect intrinsic properties of the anatase phase, independent of the sample preparation method. While oxygen defects are always present to some extent after the surface preparation, we have found that exposure to UV photons induces a much larger amount of vacancies, and provides a substantial electron doping. Hence, we could tune the electron density over more than two orders of magnitude by varying the beam intensity and the oxygen partial pressure during the ARPES experiment, and explore samples with carrier densities in the $10^{18}-10^{20}$~cm$^{-3}$ range.

\begin{figure*}
\begin{center}
\includegraphics[width=174mm]{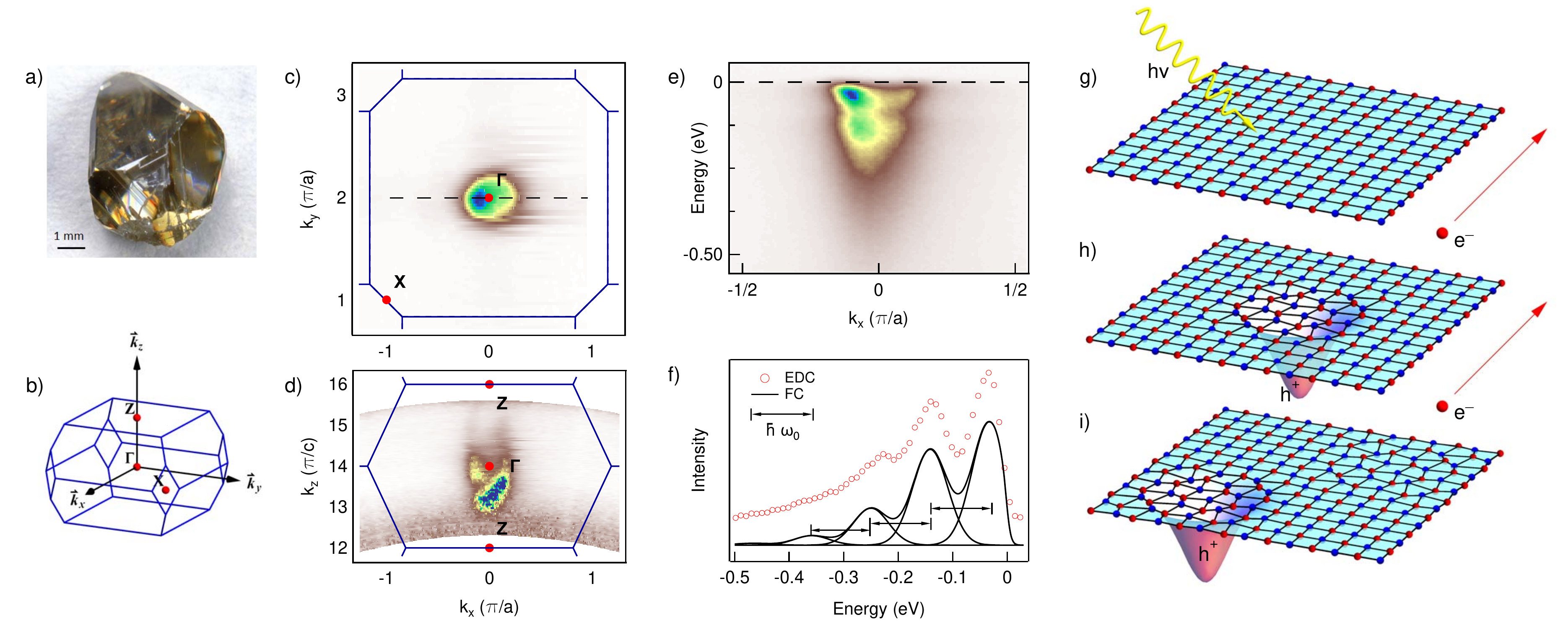}
\caption{\label{fig1} \textbf{(a)} An as-grown anatase single crystal. \textbf{(b)} The BZ of anatase.  \textbf{(c,d)} CE maps at $E_F$ (T$=20$~K, $h\nu=85$~eV) of electron-doped anatase $(001)$ in the $k_xk_y$ \textbf{(c)} and $k_xk_z$ \textbf{(d)} planes, respectively. The blue lines outline the boundaries of the 3D BZs. \textbf{(e)} $E$~\textit{vs.}~$k$ dispersion of the bottom of the conduction band for a sample with $n_e\simeq3.5\times10^{19}$~cm$^{-3}$. \textbf{(f)} ARPES intensity measured at $k=k_F$ for a sample with $n_e\simeq5\times10^{18}$~cm$^{-3}$. The solid line is a Frank-Condon line shape. Voigt peaks of width $\Delta E=90$~meV (FWHM) are separated by $108$~meV, while intensities follow a Poisson distribution. \textbf{(g-i)} Cartoon of the  polaron formation induced by the photoemission process, showing the solid in its ground state (g) and two possible final states (h,i) of ARPES.}
\end{center}
\end{figure*}

Fig.~1(c-d) shows two ARPES constant energy (CE) maps measured at the Fermi level in two perpendicular sections of momentum space, parallel (c) and perpendicular (d) to the $(001)$ surface. The two planes completely define the three-dimensional (3D) Fermi surface (FS) of an ellipsoidal electron pocket, elongated in the $k_z$ direction and centered at the $\Gamma$ point, center of the 3D Brillouin zone (BZ) (Fig.~1(b)). The data are in agreement with theoretical calculations for the bottom of the conduction band of anatase TiO$_2$ \cite{Chiodo2010}, and more importantly, they establish the existence of conduction electrons with a well defined FS in electron-doped anatase.
Note that the closed FS contour proves the 3D nature of the electronic dispersion, in contrast with the two-dimensional character of the metallic states observed at the surface of SrTiO$_3$ \cite{Santander-Syro2011,Meevasana2011} and KTaO$_3$ \cite{King2012}.

The ARPES intensity map of Fig.~1(e) illustrates the energy-momentum dispersion of the conduction states for a sample with  electron density  $n_e\sim3.5\times10^{19}$~cm$^{-3}$, as determined from the volume of the electron pocket.  It consists of a shallow QP band with minimum at $-40$~meV, which crosses  the Fermi level  $E_F$ at $k_F=\pm0.12$~{\AA}$^{-1}$. Remarkably, the QP band is followed by a satellite with a similar dispersion at $\sim$100~meV  higher binding energy, and by a broad tail. The satellite is more clearly resolved in samples with lower carrier densities, \textit{e.g.} in the spectrum of panel (f)  measured on  a sample with $n_e\sim5\times10^{18}$~cm$^{-3}$, where  also  a weaker replica is visible  $\sim$100~meV  below the first satellite.

The intensity distribution of Fig.~1(e) reveals the clearly dispersive nature of the states, but also a substantial renormalization of the spectral function, incompatible with a simple scenario of a metal with weakly interacting electrons. A parabolic fit yields an effective mass of $m_{xy}^{\ast}=(0.7\pm0.05)m_e$, where $m_e$ is the bare electron mass. A comparison with $m_{xy}=0.42m_e$ from a band structure calculation \cite{Hitosugi2008}, yields a mass renormalization $(m^{\ast}_{xy}/m_{xy})\sim1.7$. Indeed, at these carrier densities ($n_e\sim10^{18}-10^{19}$~cm$^{-3}$) the low energy states appear to fit in the intermediate regime of the so-called large polarons. The electron-phonon (e-ph) coupling induced by the ionic anatase lattice causes the QP dressing and the mass enhancement, but the polaron wavefunctions extend over several lattice constants.  We stress here that, although similar claims are valid for the rutile phase of TiO$_2$ \cite{Hendry2004}, the same ARPES experiment repeated on several (001) rutile surfaces failed to show any trace of a metallic edge, even after a long exposure to the photon beam. \\
The well-known Franck-Condon (FC) scenario for an electron coupled to a vibrational mode provides a schematic but instructive guideline to interpret the data \cite{Sawatzky1989}. In the electron-removal spectrum, the main ``zero phonon'' peak is followed by a progression of vibronic satellites, separated by the phonon energy $\hbar\omega_0$, with peak intensities following the Poisson distribution \cite{Mahan1993}. A FC line shape indeed provides a good qualitative description of the spectrum of Fig.~1(f) for a phonon energy of $\hbar\omega_0=108$~meV, the energy of a longitudinal optical (LO) $E_u$ phonon mode observed by Raman  spectroscopy \cite{Gonzalez1997}. Within the photoemission literature, the series of distinct satellites is hardly observable, and the distinctive sign of a polaronic system is the characteristic ``peak-dip-hump'' spectrum \cite{Shen2004, Mannella2005, Massee2011,Lee2013}.
Fig.~1(g-i) illustrates the physical picture underlying the polaron state as measured by ARPES.  After absorbing a photon (g), the solid is left with a photohole coherently coupled to a phonon cloud, moving through the lattice in its ground state (h), or in one of its vibrational excited states (i). The case (h) corresponds to the QP band, while (i) corresponds to the satellite replica(s).

\begin{figure}[b]
\begin{center}
\includegraphics[width=86mm]{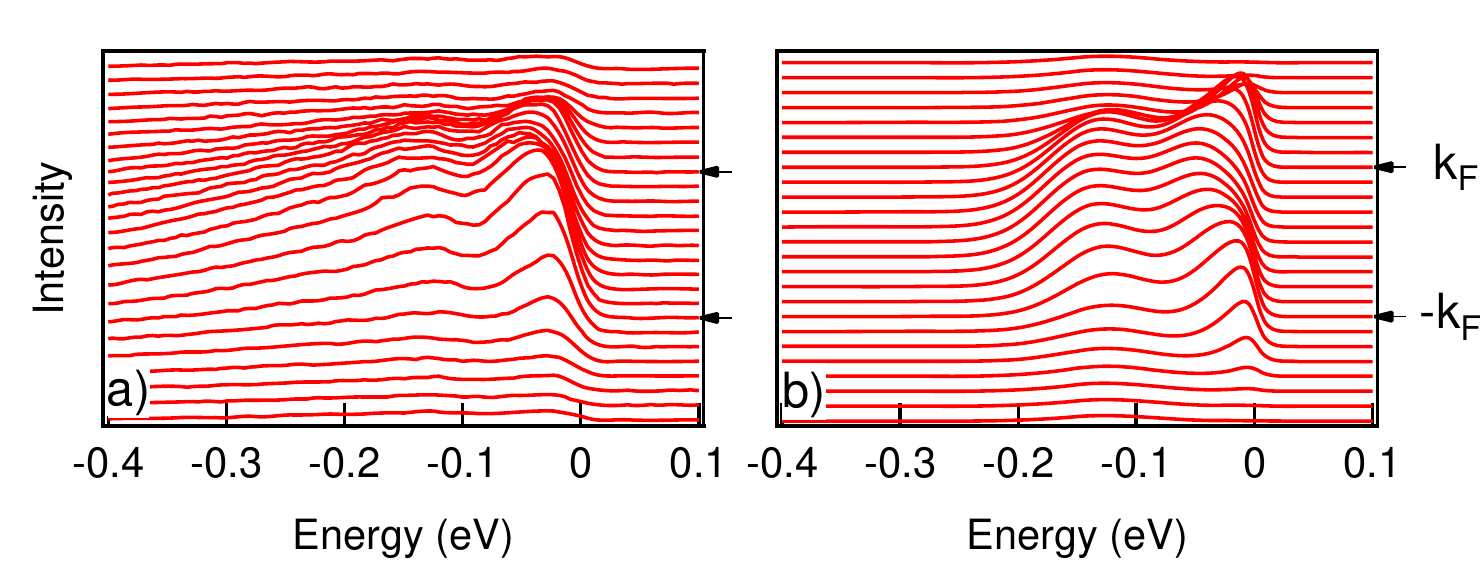}
\caption{\label{fig2}
\textbf{(a)} Spectra extracted from the intensity map of Fig.~1(e) in the range $-0.25 < k_x<0.25$~{\AA}$^{-1}$.
\textbf{(b)} Spectral function $A(k,E)$ for the Fr\"{o}hlich polaron, including the QP term and the leading $l=1$ incoherent contribution, corresponding to the first phonon satellite. }
\end{center}
\end{figure}

In order to get more quantitative information on the effect of e-ph coupling in this (unusual) regime, we consider, as appropriate for a polar material, a Fr\"{o}hlich polaron model. It is characterized by the non-local interaction between the electron and an LO phonon branch \cite{Mahan1993}:

\begin{align}\label{Frolich}
 \hat H_{\textrm{e-ph}}&=&\frac{1}{\sqrt \nu}\sum_{\vec k,\vec q}\frac{M}{|q|}c^\dagger_{\vec k+\vec q}c_{\vec k}(b_{\vec q}+b^\dagger_{-\vec q})~,\nonumber\\
M^2&=&\frac{4\pi \alpha \hbar (\hbar \omega_0)^{3/2}}{\sqrt{2 m_b}}~.
  \end{align}

\noindent $c^\dagger_{\vec k}$ and $b^\dagger_{\vec q}$ create, respectively, an electron with wave vector $\vec k$ and a phonon with wave vector $\vec q$. $m_b$ is the band mass of the uncoupled electron and the dimensionless constant $\alpha$ defines the e-ph coupling strength. For sufficiently small dopings, a single-polaron approach is appropriate, since the effective polaron-polaron interaction is weak in this limit.

The ARPES spectrum is proportional, via  dipole  matrix elements, to the one-particle spectral function $A(\vec{k},E)$.
For  a Fr\"{o}hlich   polaron it takes the form: $A(\vec{k},E)=A_{c}^{(0)}(\vec{k},E)+A_{inc}^{(l>0)}(\vec{k},E)$.
The  coherent  QP spectrum $A_{c}^{(0)}(\vec{k},E)$ corresponds to transitions  to an excited state where the lattice remains unperturbed. It follows the renormalized band dispersion
$E^{(0)}(\vec k)=\hbar^2k^2/2m^{\ast}-\mu$~, where $\mu$ is the chemical potential.
The remaining $l>0$ contributions are the  incoherent  parts of the spectrum, and involve $l$ phonon excitations in the final state.
The leading $l=1$ term can be calculated analytically. It exhibits a logarithmic singularity at energy $E^{(1)}(\vec{k})=E^{(0)}(\vec{k})-\hbar\omega_0$, \textit{i.e.}, the same dispersion as the QP band, and increased broadening  (Suppl. Inf.). The $l>1$ terms  yield a fairly homogeneous background in the energy ranges $-l\hbar\omega_0-\mu < E < -l\hbar\omega_0$.
Fig.~2 compares the experimental spectra extracted from the intensity map of Fig.~1(e) with the calculated spectral function, including the $l=0$ and $l=1$ terms, with $\hbar\omega_0=108$~meV. A Gaussian broadening ($85$~meV  full width at half maximum - FWHM) was applied to the theory. It accounts for the coupling to low-energy phonons and other scattering mechanisms, and for the instrumental resolution ($30$~meV). The overall agreement between theory and experiment is very satisfactory,  considering that the theory does not include the background tail of the $l>1$ terms.

\begin{figure}
\begin{center}
\includegraphics[width=86mm]{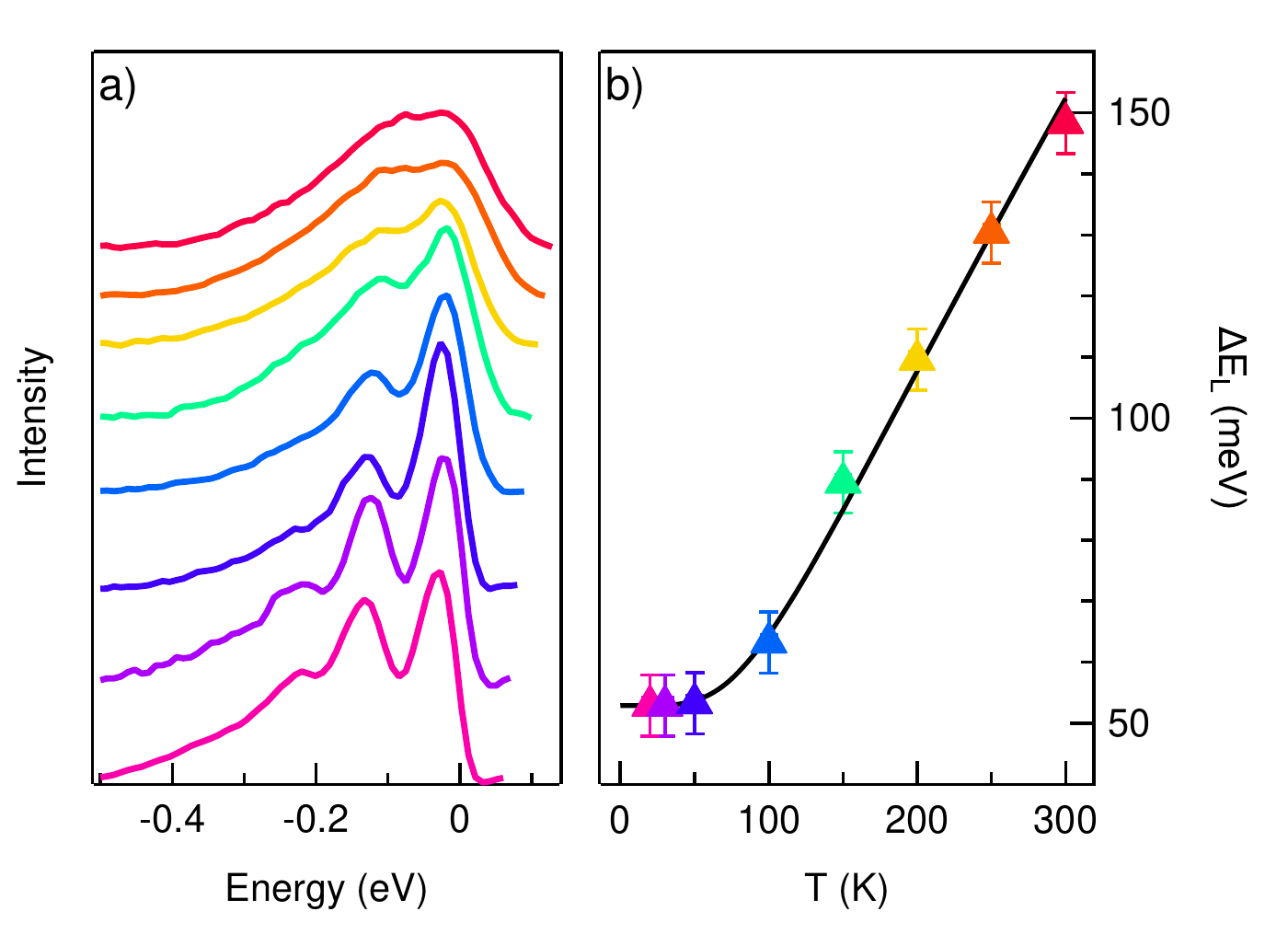}
\caption{\label{fig3} \textbf{(a)} Spectra measured between $T=20$~K and $T=300$~K on a sample with $n_e\simeq5\times10^{18}$~cm$^{-3}$, showing a progressive broadening with increasing temperature. The signal extracted from the shallow electron pocket was integrated over a momentum range of $-k_F<k_x<k_F$. \textbf{(b)} Temperature-dependent intrinsic QP line width. Colors of the symbols correspond to those of the spectra in (a). The line is a Bloch-Gr\"uneisen curve.}
\end{center}
\end{figure}

The above calculation is too simple to provide an accurate estimation on the strength of the e-ph coupling. We can assess it more reliably analyzing the intensity distribution of the ARPES signal between the two branches. Quite generally, spectral weight is
transferred from the QP band to the incoherent satellites as the e-ph coupling is increased.
The coherent fraction $Z(k_F)$ of the total spectral weight can be inferred from the experimental spectrum of Fig.~1(f), yielding $Z(k_F)=0.36$. For this value, diagrammatic quantum Monte Carlo simulations of the electron addition spectrum deduce an e-ph coupling constant $\alpha\simeq2$ \cite{Mishchenko2000}, in a regime of intermediate coupling. Using as a starting point the experimentally observed mass renormalization factor of 1.7, the same numerical calculations predict $\alpha\simeq2.5$, in fairly good agreement and in the same coupling regime.

The size of the e-ph coupling also has an influence on the energy broadening of the photoemission signal. This is illustrated by the spectra in Fig.~3(a), measured between $20$~K and $300$~K on a sample with $n_e\simeq5\times10^{18}$~cm$^{-3}$.  The spectral line shape  exhibits a  temperature dependence besides  the trivial broadening of the Fermi cutoff. The instrinsic (Lorentzian) FWHM $\Delta E_L$ of the QP peak, extracted from the spectra after removal of an experimental (Gaussian) broadening, is shown in Fig.~3(b). $\Delta E_L$ exhibits an approximately linear  $T$ dependence  above $T=150$~K, and saturates below $T=50-60$~K to a value $\sim55$~meV,  which includes contributions from impurity scattering and from the finite photoelectron  lifetime. The data are well described by  the Bloch-Gr\"uneisen curve modeling resistivity in metals \cite{ALLEN1994}. From its high temperature limit $\Delta E_L=2\pi\lambda k_BT$ one  can  extract the mass-enhancement parameter $\lambda\simeq0.7$, which again yields $(m_{xy}^{\ast}/m_{xy})=1+\lambda=1.7$.  This $\lambda$ accounts for the electron interaction with all the phonons, in particular with the low-energy acoustic modes, and should not be confused with $\alpha$ introduced above, which embodies the coupling with the single LO mode at 108~meV .

We now turn to the doping dependence of the spectra. During the ARPES measurement,  oxygen vacancies are created by the photon beam in a thin layer below the surface.  If the sample is at the same time exposed to a small O$_2$ partial pressure, a competing re-oxidation process takes place. A dynamic equilibrium is reached, which depends on the photon flux at the sample. This offers a unique opportunity to study the electronic states of anatase at various electron densities, without changing \textit{ex situ}  the stoichiometry or adding  extrinsic  impurities. In  the following experiment, we kept the photon flux on the sample constant, and varied the oxygen pressure between $2\times10^{-10}$ and $5\times10^{-8}$ mbar. The corresponding electron densities, estimated from the volume of the electron pocket at the bottom of the CB, varied between $\sim5\times10^{20}$~cm$^{-3}$ and $10^{18}$~cm$^{-3}$, respectively. At constant oxygen pressure, data could be collected for several hours without appreciable changes.

Fig.~4 shows spectra of the CB for various doping levels. Panel (a) for $n_e=5\times10^{18}$~cm$^{-3}$ shows a very shallow electron pocket and well-defined satellites.  At $T=20$~K, the momentum width $\Delta k$ of the ARPES spectral function at $k_F$ gives a QP coherence length $l=1/\Delta k=7$~{\AA}. Therefore, in this low-density limit the polaronic QPs cannot move freely for more than $\sim2$ unit cells.  Increasing the carrier density to $n_e=3\times10^{19}$~cm$^{-3}$ in (b) and to $n_e=1\times10^{20}$~cm$^{-3}$ in (c), one observes that : i) the QP band dispersion deviates from a parabola, and  ii) near the bottom of the band, the QP intensity is reduced and the satellite spreads into the background. At $n_e=3.5\times10^{20}$~cm$^{-3}$ the QP band is visible only near the Fermi level crossings at $\pm k_F$.  For smaller wave vectors, the QP intensity is strongly suppressed, with most of the spectral weight spread over a broad energy range  (the ``hump'').

\begin{figure}
\begin{center}
\includegraphics[width=86mm]{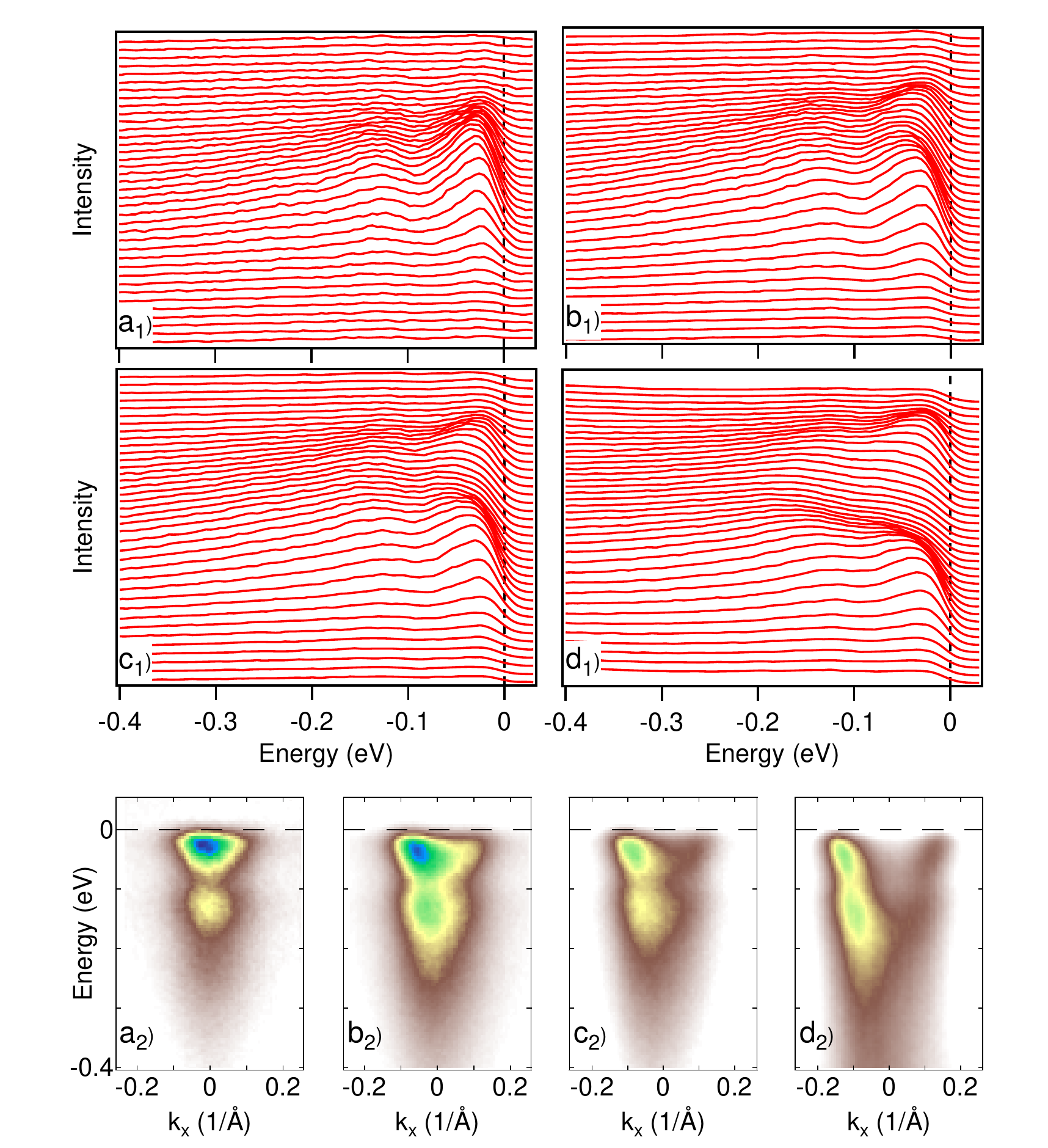}
\caption{\label{fig4}Evolution of the CB states with doping. The ARPES spectra ($h\nu=85$~eV) were measured at $T=20$~K along the dashed line ($k_y=2\pi/a$) of Fig.~1(c) for  samples with $n_e=5\times10^{18}$~cm$^{-3}$ {\bf(a1)}, $n_e=3\times10^{19}$~cm$^{-3}$ {\bf(b1)}, $n_e=1\times10^{20}$~cm$^{-3}$ {\bf(c1)}, and $n_e=3.5\times10^{20}$~cm$^{-3}$ {\bf(d1)}. The same data are also presented as $E$~\textit{vs.}~$k_x$ image plots in the bottom panels {\bf(a2-d2)}. A doping scan with finer steps is presented as a movie attached to the suppl. info.}
\end{center}
\end{figure}

The dispersion of Fig.~4(d) in the vicinity of $E=\hbar\omega_0$ is reminescent of the characteristic kink structure encountered in several Fermi liquids with more moderate e-ph coupling. The evolution of the CB states with doping  indeed  suggests a breakdown of the single-polaron  picture, as charges added to the CB progressively screen the e-ph interaction. Polarons, which are well defined QPs at low density, eventually  lose coherence and dissociate into an electron liquid  coupled to the phonon(s). The spectral weight distribution for high dopings can be reproduced by standard perturbation theory \cite{Mahan1993} (Suppl. Inf.).

Our observations shed light on the conduction mechanisms taking place in anatase-based devices, and in particular on the role of the e-ph coupling, which has been shown to represent the dominant scattering process at typical operating temperatures in pristine and doped films \cite{Furubayashi2005}. The tunability of the doping level by UV (or e-beam) illumination over a very broad range is attractive for numerous applications, and in particular in the field of transparent conductors. Carrier densities  $>10^{20}$~cm$^{-3}$, necessary for thin film operation, can be reached without extrinsic metal dopants, which are additional scattering centers \cite{Gordon2000}. Moreover, the possibility of after-growth patterning of conductive paths could yield important practical advantages. Namely, the initial growth conditions could be set independently of the required final conductivity, and etching processes often involved in oxide structuring could potentially be avoided.

Finally, the present study is likewise relevant for anatase-based devices employing nanostructured materials, where the overall transport properties depend on inter- as well as intra-particle processes. The crossover from a polaronic to a diffusive regime is expected to occur when the overlap between the polaron clouds becomes significant. The ARPES data of Fig. 4 suggest that this happens around $n_e^\ast\simeq 10^{19}$~cm$^{-3}$. Estimating the polaron radius $r_p$ from the average separation between polarons $d\sim n^{-1/3}= 2r_p$, gives $r_p\sim 20$~{\AA}. By comparison, Fr\"{o}hlich's model in the intermediate coupling regime yields $r_p=\sqrt{\hbar/2 m^* \omega_0}\simeq10$~{\AA}. Both values are much smaller than the typical dimensions (few nm) of anatase nanoparticles considered for applications. Therefore, the polaronic nature of the QPs and their evolution upon electron doping  must necessarily be taken into account when modeling transport in actual devices.

We acknowledge support by the Swiss NSF, namely through Grant N PA00P21-36420 (L.M.). We thank N. Mannella, A. Petrozza, C. Tournier-Colletta and A. Crepaldi for discussions. The Advanced Light Source is supported by the Director, Office of Science, Office of Basic Energy Sciences, of the U.S. Department of Energy under Contract No. DE-AC02-05CH11231.

\end{document}